\documentclass[12pt]{article}
\begin{document}
\title{The ``Thermodynamic'' Universe}
\author{B.G. Sidharth\\
International Institute for Applicable Mathematics \& Information Sciences\\
Hyderabad (India) \& Udine (Italy)\\
B.M. Birla Science Centre, Adarsh Nagar, Hyderabad - 500 063 (India)}
\date{}
\maketitle
\begin{abstract}
Much of the twentieth century physics has considered rigid laws-- even Quantum Mechanics and Statistical Mechanics were based on such laws. These laws operated in a smooth spacetime manifold. However more recent approaches as in Quantum Gravity schemes or Quantum Super String theory deal with a non differentiable spacetime manifold. With this we have to consider the new paradigm of Dark Energy of Zero Point Field. The picture that now emerges is that of a universe that is ``Thermodynamic''. We investigate how this can explain phenomenon like the velocity of light and also provide tests for experimental verification.
\end{abstract}
\section{Introduction}
The Laws of Physics were considered to be rigid laws. With the advent of Statistical Mechanics and Thermodynamics, a new line of thinking based on a collection of a huge number of particles or ensembles came into being. Now there were probabilities, even though the underlying microphysical laws were the fixed laws of physics. This lead to, what is sometimes referred to as the Boltzmann paradox-- how can there be irreversibility of time at the macro scale, even though the underlying laws were the time reversible laws of physics? It was only in the last century that it was realized that the earlier ideas were based on what have been called integrable systems, which had been first considered by Poincare. In these systems, it was possible to eliminate or include interactions with a suitable set of generalized coordinates obtained by a suitable canonical transformation. This would correspond to the free particle approximation in Statistical Mechanics.\\
However such systems were the exception rather than the rule as previously supposed. In other words most of the real world systems were non integral-- and this would explain the paradox of time irreversibility \cite{prig}.\\
Though Quantum theory is based on the Heisenberg Uncertainty Principle, it must be emphasized that the underlying equations are once again time reversible, be they the Schrodinger equation or the Klein-Gordon and Dirac equations. Moreover the spacetime of Quantum theory, including Quantum Field Theory has been either Newtonian or Lorentzian, both of which define a differentiable spacetime manifold.\\
More recently, such a differentiable space time manifold has been abandoned in approaches like Quantum Super Strings or those of Quantum Gravity. This has reslted in progress towards a unification of gravitation with other fundamental interactions. In this picture, spacetime is a violent ``foam''. As Wheeler put, ``No prediction of space time, therefore no meaning of space time is the verdict of the Quantum Principle. That object which is central to all of Classical General Relaivity, the four dimensional space time geometry, simply does not exist, except in a Classical approximation.''\\
Another new element that has been added to the scenario, in the past few years is that of an underlying Dark Energy or in older language the Zero Point Field. This has been confirmed in the past few years thanks to the work of Perlmutter, Schmidt, Kirshner and others as also the Wilkinson Microwave Anisotropy Probe and the Sloan Digital Sky Survey (Cf.refs.\cite{cu,uof,perl} and several references therein). Infact this was the breakthrough of the year 2003, of the American Association for Advancement of Science \cite{science}. Dark Energy or the Zero Point fluctuations endue the universe with a stochastic or Brownian character, which again finds expression in a non commutative geometry \cite{bgsfpl}.
\section{The Velocity of Light}
Given the above scenario, the question then arises, how can we explain the constancy of the velocity of light? For this we first observe that the inertial mass itself can be considered to be arising from the Zero Point Field, as demonstrated some years ago by the author and also Rueda, Hirsch and Puthoff \cite{heap, rueda,herch}. In this scenario inertia is a result of the viscous resistence of the Zero Point Field. So if there is no inertial mass, then there is no resistence and the velocity should become infinite, or in practical terms, very large.\\
Infact if we start with the Langevin equation in a viscous medium \cite{rief,balescu} then as the viscosity becomes vanishingly small, it turns out that the Brownian particle moves according to Newton's first law as if there were no force acting on it, that is with a constant velocity. Moreover this constant velocity is given by (Cf.refs.\cite{rief,balescu}),
\begin{equation}
c^2 \equiv \langle v^2 \rangle = \frac{kT}{m}\label{e1}
\end{equation}
Let us consider (\ref{e1}) with minimal values of $T$ and $m$, in the real world. Infact we have the thermodynamic Beckenstein formula \cite{ruffini}
\begin{equation}
T = \frac{\hbar c^3}{8\pi k GM}\label{e2}
\end{equation}
We consider in (\ref{e2}) the entire universe so that the mass $M$ is $\sim 10^{55}gms$. Substitution in (\ref{e2}) gives
\begin{equation}
T = \frac{10^4}{10^{32}} \sim 10^{-28^{\circ K}}\label{e3} 
\end{equation}
We next consider in (\ref{e1}),m to be the smallest possible mass. From thermodynamical considerations, Landsberg has shown that this is given by \cite{land}
\begin{equation}
m \sim 10^{-65} gms\label{e4}
\end{equation}
The same equation (\ref{e4}) can also be obtained from a different point of view, namely the Planck scale underpinning for the universe in modern Quantum Gravity approaches \cite{fpl}. Substitution of (\ref{e3}) and (\ref{e4}) in (\ref{e1}) gives
$$
\langle v^2 \rangle = \frac{kT}{m} = \frac{10^{-16} \times 10^{-28}}{10^{-65}} = 10^{21}, i.e.$$
\begin{equation}
v = c\label{e5}
\end{equation}
We can see from (\ref{e1}) and (\ref{e5}) that the velocity $c$ is exactly the velocity of light!\\
The question is, for how long such a particle with vanishingly small inertial mass can maintain the constant velocity $c$, that is the velocity of light. Infact the Compton time for a particle with mass given by (\ref{e4}) as can be easily checked is, $10^{17}secs$, which is the age of the universe! Alternatively let us, in the spirit of the Beckenstein formula, (\ref{e2}) compute the Beckenstein life time from the well known formula 
\begin{equation}
t = \Theta^3 \pi e^4 / \hbar c^4 m\label{e6}
\end{equation}
Substitution in (\ref{e6}) gives us again, the age of the universe.
\section{The Zero Point Field and Non Commutative Geometry}
In the more recent approaches of high energy physics, the smooth space time manifold has been replaced by a non differentiable space time. This is linked to a minimum interval, typically the Planck scales, and as is well known implies also a non commutative space time \cite{r1,r2,r3,r4}. Such a non commutative space time also implies a breakdown of Lorentz symmetry at very high energies-- this is a test which will hopefully be carried out on ultra high energy cosmic rays in the coming years (Cf. ref.\cite{ijtp} and references therein). We will now discuss another simple test for the non commutative nature of space time.
\section{The Test}
We first observe that non commutativity means that simple coordinates like $x$ and $y$ do not commute but rather we have relations of the type 
\begin{equation}
[x,y] = \beta \Theta\label{e1a}
\end{equation}
where $\beta \sim l^2 , l$ being the minimum extension and $\Theta$ are matrices. The Equation (\ref{e1a}) suggests that the coordinates also contain some type of a momentum with a suitable dimensional factor \cite{bgsnc,ffp4}:
\begin{equation}
y = h' p_x\label{e2a}
\end{equation}
It has been shown that
\begin{equation}
h' = l^2/h\label{e3a}
\end{equation}
On the other hand, it has been shown by the author, Saito and others \cite{bgsncb,saito} that the above non commutative geometry gives rise to a magnetic field $\vec{B}$ given by
\begin{equation}
B l^2 = hc/e\label{e4a}
\end{equation}
where $B = |\vec{B}|$. Use of (\ref{e4a}) with (\ref{e2a}) and (\ref{e3a}) gives
\begin{equation}
y = \frac{c}{eB} p_x\label{e5a}
\end{equation}
Equation (\ref{e5a}) is familiar from the theory of a particle in a uniform magnetic field, first worked out by Landau \cite{landau,greiner}. What happens there is, given a uniform magnetic field along the $z$ axis, the particle in the classical sense executes circles in the $x,y$ plane with quantized energy levels given by, with the usual notation 
\begin{equation}
E = (n + \frac{1}{2}) h \omega_B + p_z^2/ 2m - \mu \sigma B/s\label{e6a}
\end{equation}
where $\omega_B$ is given by
$$\omega_B = |e|B/mc$$
These are the so called Landau levels. Thus the non commutative geometry would show up as a quantization of the energy as in (\ref{e6a}). It must be observed that in (\ref{e6a}) $p_z$ is itself not quantized.

\end{document}